\documentclass[aps,prl, twocolumn, superscriptaddress]{revtex4-1}
\usepackage{graphicx}
\usepackage{amsmath}
\usepackage{amsfonts}
\usepackage{amsthm}
\usepackage{amssymb}
\usepackage{amsbsy}
\usepackage{wasysym}
\usepackage{bm}
\usepackage{mathrsfs}
\usepackage{color}
\usepackage{times}
\usepackage[resetlabels]{multibib}

\newcites{sec}{Supplementary References}

\newcommand{\beq}{\begin{equation}}
\newcommand{\eeq}{\end{equation}}
\newcommand{\bea}{\begin{eqnarray}}
\newcommand{\eea}{\end{eqnarray}}

\newlength{\myL}

\newcommand\hexp{}

\def\ket#1{{\left|#1\right\rangle}}

\def\bk{\mathbf{k}}
\def\bq{\mathbf{q}}

\def\bR{\mathbf{R}}

\def\be{\begin{eqnarray}}
\def\ee{\end{eqnarray}}

\begin{document}

\title{Wannier Permanent Wave Functions and Featureless Bosonic Mott Insulators on the 1/3
filled Kagome Lattice
}\author {S.A. Parameswaran}
\affiliation{Department of Physics, University of California, Berkeley, CA 94720, USA}
\author{Itamar Kimchi}
\affiliation{Department of Physics, University of California, Berkeley, CA 94720, USA}
\author{Ari M. Turner}
\affiliation{Institute for Theoretical Physics, University of Amsterdam, Science Park 904, P.O. Box 94485, 1090 GL Amsterdam, The Netherlands}
\author{D. M. Stamper-Kurn}
\affiliation{Department of Physics, University of California, Berkeley, CA 94720, USA}
\author{Ashvin Vishwanath}
\affiliation{Department of Physics, University of California, Berkeley, CA 94720, USA}

\date{\today}
\begin{abstract} We study  Bose-Hubbard models on tight-binding, non-Bravais lattices, with a filling of one boson per unit cell -- and thus fractional {\it site} filling. We discuss situations where no classical bosonic insulator, which is a product state of particles on independent sites, is admitted. Nevertheless we show that it is possible to construct a quantum Mott insulator of bosons, if a trivial band insulator of {\em fermions} is possible at the same filling. The ground state is simply a permanent of exponentially localized Wannier orbitals. Such a Wannier permanent wave function is featureless in that it respects all lattice symmetries, and is the  unique ground state of a parent Hamiltonian that we construct.
Motivated by the recent experimental demonstration of a kagome optical lattice of bosons, we study this lattice at 1/3 site filling. Previous approaches to this problem have invariably produced either broken-symmetry states or topological order. Surprisingly, we demonstrate that a featureless insulator is a possible alternative and is the exact ground state of a local Hamiltonian. We briefly comment on the experimental relevance of our results to ultracold atoms as well as to $1/3$ magnetization plateaus for kagome spin models in an applied field.

\end{abstract}
\maketitle
\noindent{\it Introduction.---} Much recent activity in condensed matter physics has focused on identifying nontrivial phases of matter which are inherently quantum-mechanical and cannot be understood by perturbing around  a straightforward classical limit.
Proposed systems where such phases might occur include strongly correlated electronic systems, frustrated antiferromagnets, and insulating phases of bosonic lattice systems, the last of which is our focus in this paper.
Unlike fermions, bosons are precluded from forming noninteracting band insulators so {\it all} crystalline insulators require interactions: hence the term `Mott insulators' \cite{Fisher:1989p1}. Mott insulators realized at integer fillings of bosons per site have a classical description deep within the insulating state in terms of a fixed integer number of particles per site, which means that to find nontrivial states, fractional site filling is desirable.

Theoretically, it has been  proven \cite{Lieb:1961p1, Hastings:2004p1} that insulators at  fractional filling per {\it unit cell}  cannot be  `featureless': they form either crystals with an enlarged unit cell by breaking lattice translation symmetry, or exotic phases with topological order - i.e. phases with emergent excitations that carry  unusual statistics.
 Experiments on cold atoms in optical lattices have extensively explored  Mott insulating phases on a variety of simple lattices \cite{Greiner:2002p1, foll05noise, foll06shell, camp06shell}. All these are Bravais lattices, with one site per unit cell -- so that the site and unit cell fillings are identical --  ruling out featureless states at fractional site filling.

 However, recently, more complicated optical lattices with a basis -- such as the honeycomb \cite{Tarruell:2012p1} and kagome \cite{Jo:2012p1} structures  -- have been created, and the Mott and superfluid states in them have begun to be studied. This naturally  leads us to consider  fractional {\it site} filling, but integer unit cell filling. Are  symmetry-breaking or topological order still the only alternatives? We will study examples where the answer is {\it no}, but yet quantum fluctuations of bosons must be significant, even deep within a featureless insulating phase.

\begin{figure}
\includegraphics[width=.9\columnwidth]{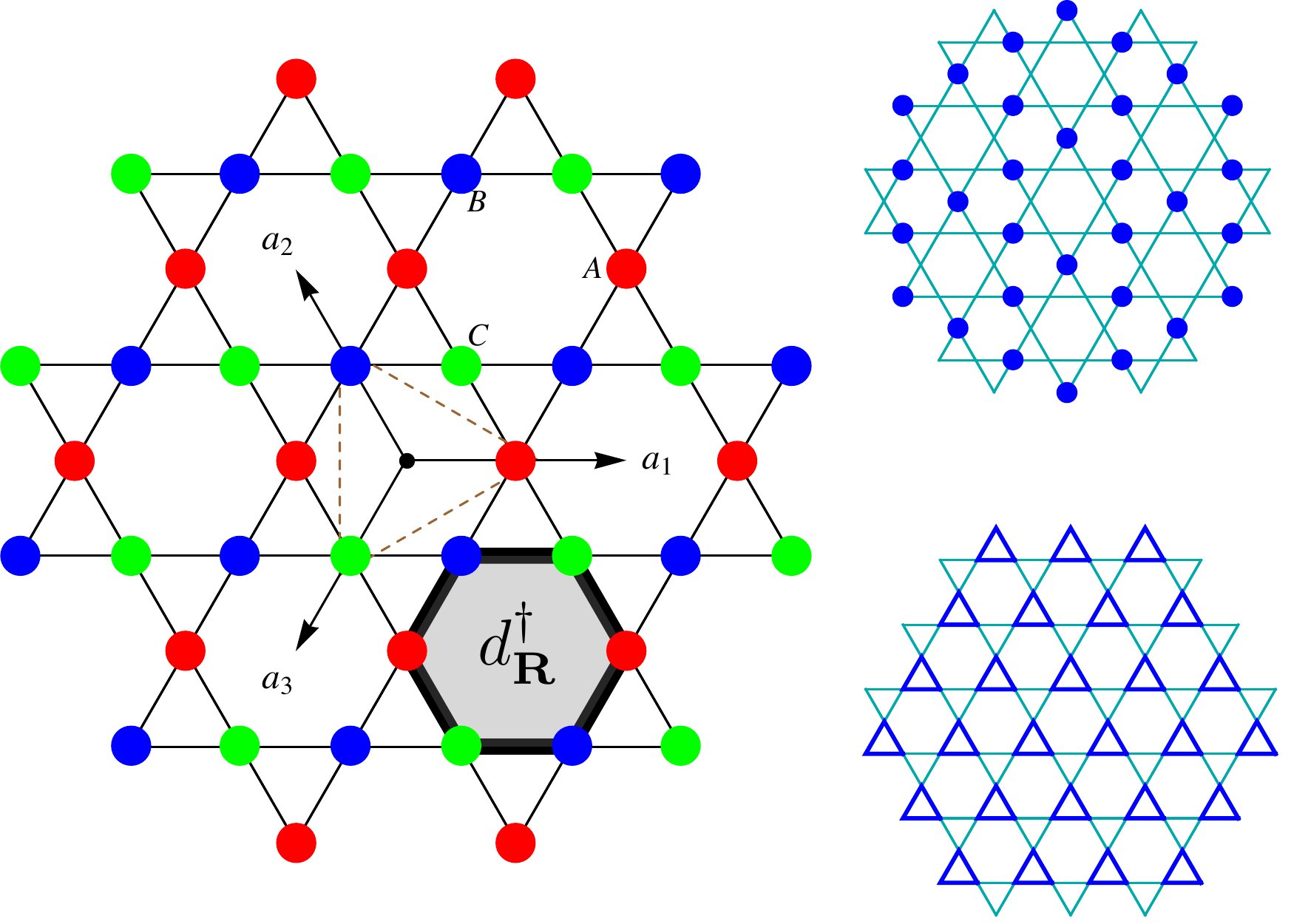}
\caption{\label{fig:kagome1}(a) Kagome lattice and notation used in the text.  Standard methods to construct featureless phases break symmetries, e.g. (b)  `$\sqrt{3}\times\sqrt{3}$' state  and (c)  triangular `molecular orbital' state.}
\end{figure}

As an example of why a featureless Mott phase at fractional site filling can be counterintuitive, consider the kagome lattice at a filling of one boson per unit cell, or $1/3$ site filling. Symmetries conflict with the usual caricature of a Mott  wavefunction: the essentially classical picture of a fixed number of bosons tied rigidly to each site.  Attempting to draw such a classical cartoon on the kagome at the given filling leads inevitably to symmetry breaking: for example, a uniform  ($\bq=0$) state that distinguishes a single site in each triangle, or the `$\sqrt{3}\times\sqrt{3}$' order which enlarges the unit cell (Fig. \ref{fig:kagome1}).
Arbitrarily choosing a unit cell, say an upward-facing triangle, and delocalizing each boson across this choice of sites gives a more quantum-mechanical insulator, but leads to a  state which  breaks point group symmetries -- specifically in this `triangular' state, that of rotation by $180^\circ$\cite{KagomeTrimer} (Fig. \ref{fig:kagome1}). Since the unit cell has three sites, which in the tight-binding limit are indivisible, one cannot write a state as a product of disjoint `molecules' while respecting the full sixfold point group symmetry. Finally, more sophisticated approaches that first implement a duality transformation and condense vortices of the dual order parameter also break symmetry at this filling \cite{Sengupta:2006p1}. Thus, for the kagome (and similar examples) fractional site filling ensures that there is no smooth connection to a trivial insulator. The usual strategies for constructing insulating phases {\it always} lead to broken symmetry, and to date no featureless insulating state has been found.

Here, we discuss a general scheme to write down wavefunctions and parent Hamiltonians for a class of such featureless quantum Mott insulators, including the $1/3$ filled kagome as our primary example. Although we are interested in the problem of bosonic Mott insulators, we
first
study the reference problem of fermions at the {\em same} filling on the {\em same} lattice. {\it If a band insulator of fermions exists, for which exponentially localized Wannier orbitals (WOs) that respect lattice symmetries can be constructed, then we show that a bosonic Mott insulating state is also feasible.}
The Mott insulating wavefunctions we construct are permanents \footnote{The permanent of a square matrix $A_{ij}$ is defined as $\text{perm}(A) = \sum_{\sigma\in S_n} \prod_{i=1}^n A_{i, \sigma(i)}$, where the sum extends over all elements $\sigma$ of the symmetric group $S_n$, i.e. it is analogous to a determinant without an alternating sign.} of these WOs-- hence, we dub them Wannier permanent wave functions-- analogous to fermionic band insulators which are determinants of the same orbitals. Furthermore, we demonstrate that these Wannier permanent wave functions are exact ground states of symmetric, local Hamiltonians of the Bose-Hubbard type. As an added bonus, the Bose-Hubbard models we discuss can also be viewed as describing XXZ spin systems with fixed magnetization (e.g., in an applied field), for which our wave functions describe fractional magnetization plateaus which do not enlarge the unit cell.  On the $1/3$ filled kagome lattice, we first find a tight-binding model which at this filling describes a band insulator, and use this to construct a  Wannier permanent wavefunction for the insulating phase. To the best of our knowledge, this is the first proposal of a featureless ground state for this problem.

Identifying a lattice like the kagome is subtle from a symmetry perspective: it entails taking the (experimentally well-motivated) tight-binding limit. Absent this restriction, the kagome, honeycomb, and triangular lattices are indistinguishable as they all have the same space group. Formally, the tight-binding limit identifies a particular {\it representation} of the symmetry, which contains more information than the space group. We can understand this intuitively: the kagome is a non-Bravais lattice with three sites per unit cell.  Although solving a tight-binding model entails a specific choice of inequivalent sites to form the crystal basis (which breaks lattice symmetry), the final many-body band
insulating wavefunction for fermions is independent of this choice and
respects symmetries. In contrast, choosing orthogonal real-space
orbitals to build a bosonic permanent is more sensitive: e.g. up- and
down-triangle permanents are distinct, and both break $180^\circ$ rotation symmetry.
 Thus, restrictions placed by tight-binding and the symmetrization of many-body bosonic wavefunctions makes constructing featureless Bose insulators challenging, in contrast to the more usual case of fermionic band insulators.

The wavefunctions we study are in a sense bosonic analogs of the Affleck-Kennedy-Lieb-Tasaki (AKLT) states \cite{Affleck87, affleck1988vbg} of spin systems, which are quantum-mechanical paramagnets. As noted in
 \cite{YaoKivelson}, the AKLT insulators are `fragile' Mott insulators of electrons, which cannot be adiabatically connected to band insulators without breaking a crystalline point-group symmetry. Similarly, our wavefunctions cannot be deformed into disjoint product states without breaking translational or point-group symmetry. However, since we discuss bosons, in contrast to \cite{YaoKivelson} we cannot consider a noninteracting limit whatever the symmetry. We note previous work \cite{Huber:2010p1} that used power-law localized WOs to study Bose condensation and charge-density-wave order in the kagome flat band is unrelated to the featureless insulators we discuss.

\noindent{\it Fermionic band insulators as bosonic Mott insulators.---}  Given a set of symmetric, exponentially localized WOs
$g_\bR(i)$, with $\bR$ a Bravais lattice vector and $i$ a lattice site,
the Wannier permanent wavefunction for a bosonic Mott insulator takes the form $\ket{\Psi_W} =\prod_\bR w^\dagger_{\bR} \ket{0}$, where $w^\dagger_{\bR} \equiv \sum_{i} g_\bR(i) b^\dagger_i$ adds a boson to the WO at $\bR$.  $\ket{\Psi_W}$ is the exact ground state of a parent Hamiltonian given by an interaction between densities projected into the WO basis.  The correlation function can be expressed as a projector over the occupied band,
\be
\langle b^\dagger_i b_j\rangle_{\Psi_W} = \sum_{\bR} g^*_\bR(i) g_\bR(j) = \mathcal{P}_\text{occ.}(j,i).
\ee
and decays exponentially, $\langle b^\dagger_i b_j\rangle \sim e^{-|i-j|/\xi}$, with $\xi$ related to the localization length of the WOs. Higher-body correlators can be similarly computed, and also decay exponentially.

WOs are constructed by integrating Bloch wavefunctions over the Brillouin zone (BZ), with an arbitrary phase choice $\varphi_{\bk}$ at each wavevector $\bk$ \footnote{Unitary transformations involving more than one band are allowed but unimportant here.}.
For fermions, the fact that the different WOs are related to each other and to the single-particle Bloch  eigenstates by unitary transformations means that  `filling'  any such complete set of single-particle states in the lowest band and then antisymmetrizing yields the {\it same} insulating wavefunction, as the resulting Slater determinants are all unitarily invariant. However, symmetrized permanents such as $\ket{\Psi_W}$ are {\it not} unitarily equivalent for different choices of $g_{\bR}(i)$, making the choice of single-particle states to `fill' crucial:  permanents for featureless Bose insulators must be built from {\it localized} and {\it symmetric} single-particle states. 

To preserve symmetry, we must be able to fix the choice of localized WO unambiguously. For a simple band  -- one that touches no other bands -- respecting time reversal and inversion symmetry\footnote{Note, just requiring inversion symmetry forces zero Chern number, so there is no {\it topological} obstruction to constructing WOs.} fixes $\varphi_{\bk}$ up to an overall sign, which is fixed by requiring continuity of $\varphi_\bk$ across  the BZ, a necessary condition for exponential localization \cite{desCloizeaux:1963p1,desCloizeaux:1964p1, desCloizeaux:1964p2, Kohn:1973p1}. Whenever a fermionic band insulator exists at a given filling, we can  find such a set of WOs giving a symmetric insulating ground state and corresponding local parent Hamiltonian-- a significant and hitherto unnoticed aspect of bosonic Mott physics.

\begin{figure}
\includegraphics[width=.9\columnwidth]{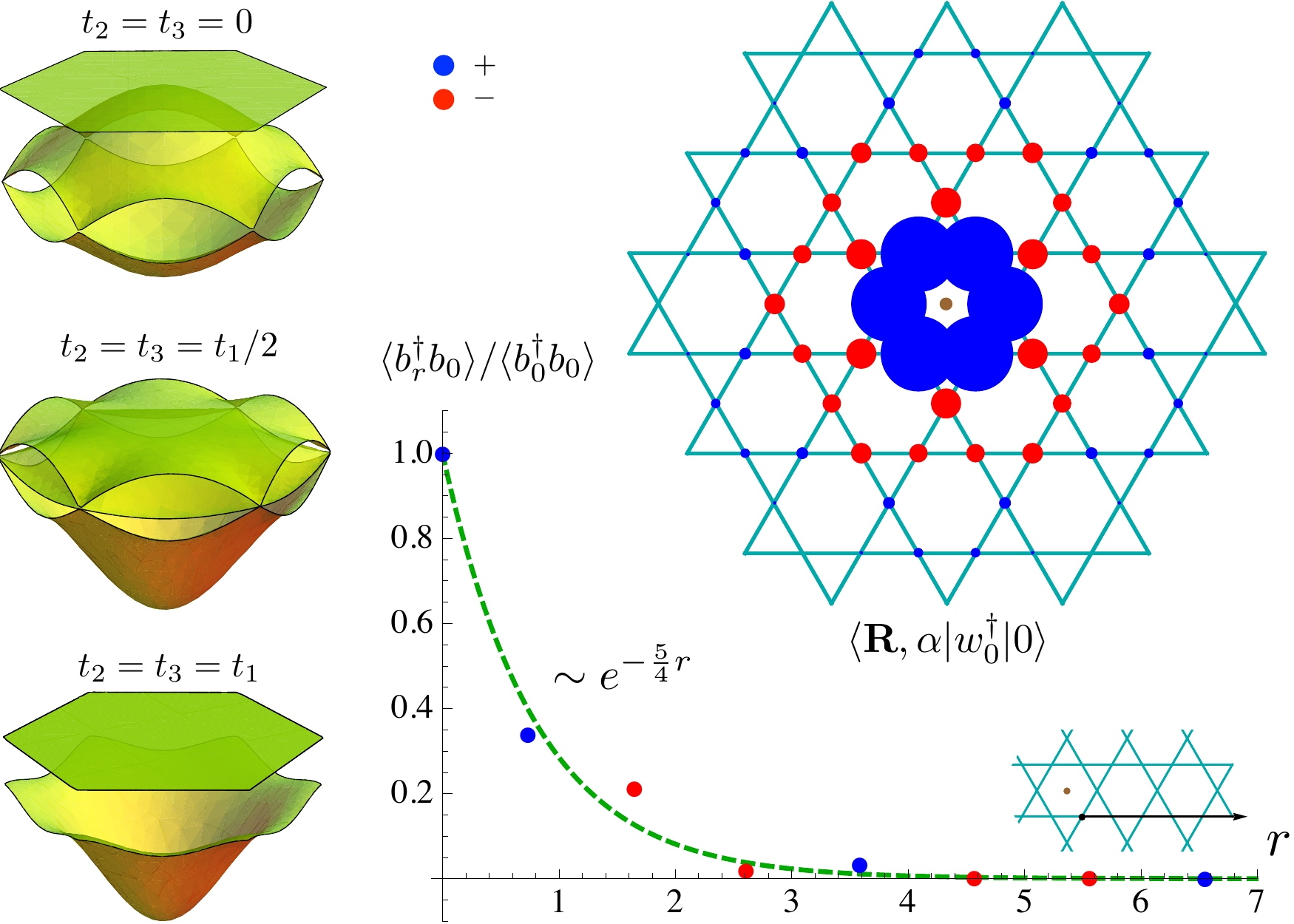}
\caption{\label{fig:WFandstates}(a) Evolution of tight-binding band structure of $H_0$; for nearest-neighbor hopping (top) all bands touch, while at the special point $t_1=t_2=t_3$ (bottom)  the lowest band is isolated and  is especially tractable. (b) WO plotted on the lattice. Size of the circles give $|g_{\bR}(i)|$; blue (red)  depict $g_\bR(i) >0$ ($<0$). Exponential decay of the boson correlator $\langle b^\dagger_r b_0\rangle$ is also shown below the WO.}
\end{figure}

\noindent{\it Kagome Lattice.---} To use this approach for the kagome lattice at $1/3$ filling we must find a fermionic band insulator (separated lowest band) on the kagome. We consider the general case $H_0[\{t_{ij}\}] \equiv -\sum_{i,j} t_{ij}  b^\dagger_i b_j$ on the kagome lattice, where the $t_{ij}$ are assumed to respect all the symmetries of the lattice. $H_0$  has three bands, and with only nearest-neighbor hopping ($t_1$) the two lower bands form gapless Dirac points (Fig.~\ref{fig:WFandstates}). Mass terms that can gap out the Dirac points break inversion or three-fold lattice symmetry (`triangular' and `$\sqrt{3}\times\sqrt{3}$' states, Fig.~\ref{fig:kagome1}). However,
as second- and third-neighbor hoppings ($t_2, t_3$) are increased from zero, the bands become degenerate at $K, K'$  when $t_2=t_3 =\frac{1}{2}t_1$; thereafter the lowest band is simple with the twofold $K$-point degeneracy transferred to the upper two bands (Fig. \ref{fig:WFandstates}. For a symmetry analysis see \cite{SupplementaryMaterial}).
This procedure preserves lattice symmetry, providing symmetric WOs for the lowest band.

\noindent{\it Wannier Permanents.---}  For clarity, we focus initially on a fine-tuned point \cite{BergmanWuBalents, BalentsFisherGirvin}, where  $t_1=t_2=t_3=t/6$, and the onsite potential $t_{ii} = t/3$; our results persist away from this point as long as the lowest band remains simple. We may rewrite $H_0$ as $
 H_0=  -t\sum_{\bR} d^\dagger_{\hexp\bR} d_{\hexp\bR}
$
by defining operators that  `smear' a boson over a single hexagon (Fig. \ref{fig:kagome1} depicts $d^\dagger_{\bR}$ for $\bR = -\mathbf{a}_2$)
 \be
 d^\dagger_{\hexp\bR} \equiv \frac{1}{\sqrt{6}}\sum_{\alpha=1,2,3} \left(b^\dagger_{\bR,\alpha} + b^\dagger_{\bR -\mathbf{a}_\alpha, \alpha}\right)
 \ee
Here, $\bR = m \mathbf{a}_1 + n\mathbf{a}_2$ lies on the triangular Bravais lattice, $\mathbf{a}_1$, $\mathbf{a}_2$ and $\mathbf{a}_3 \equiv -(\mathbf{a}_1+\mathbf{a}_2)$ are shown in Fig.~\ref{fig:kagome1}, and  $\alpha=1,2,3$ labels inequivalent sites within a unit cell, so that a kagome site $i \equiv(\bR,\alpha)$.
Since  $[d_{\hexp\bR}, d^\dagger_{\hexp\bR'}] \neq \delta_{\bR,\bR'}$, the  $d^\dagger_{\hexp\bR}$ operators are not canonical bosons and   we cannot form a Fock space out of eigenstates of $d^\dagger_{\hexp\bR} d_{\hexp\bR}$. However  in momentum space,
$ H_0 = -t{\int_{BZ}}\frac{d^2q}{(2\pi)^2} d^\dagger_{\hexp\bq}d_{\hexp\bq}
 $, where  $d^\dagger_{\hexp \bq} = N_s^{-1/2}\sum_{\bR} e^{-i\bq\cdot\bR}  d^\dagger_{\hexp \bR}
 $ with $N_s$ the number of sites.
 By a straightforward computation, $[d_{\hexp\bq},  d^\dagger_{\hexp\bq'}] = \epsilon(\bq) \delta_{\bq,\bq'}$
  with  $\epsilon(\bq) =  1 +\frac{1}{3}\sum_{i=1}^3 \cos(\bq\cdot\mathbf{a}_i)$.  Since everywhere in the BZ $\epsilon(\bq) \neq 0$, its inverse is nonsingular and we may define
$\tilde{{d}}^\dagger_{\hexp\bq} \equiv[{{\epsilon(\bq)}}]^{-1/2}  d^\dagger_{\hexp \bq}$, so that
 $H _0= -t  {\int_{BZ}}\frac{d^2q}{(2\pi)^2}\epsilon(\bq) \tilde{{d}}^\dagger_{\hexp\bq}  \tilde{{d}}_{\hexp\bq}
$, now in terms of canonical bosons $[\tilde{d}_\bq,\tilde{d}_{\bq'}^\dagger] = \delta_{\bq,\bq'}$. $H_0$ has a single band of dispersing states created by $\tilde{d}^\dagger_{\hexp\bq}$ with energy $-t\epsilon(\bq)$  that is gapped away from $E=0$.  In writing $H_0$ in this form, we have chosen a specific superposition of states in a unit cell, and therefore a specific Bloch band. The two orthogonal states in the remaining  Bloch bands form a pair of zero-energy flat bands (Fig. \ref{fig:WFandstates}). $H_0$ respects all the kagome lattice symmetries  and its lowest band is manifestly simple, as discussed above.

A single boson is placed in the WO at $\bR$ by the operator
\be\label{eq:wannierdef}
w^\dagger_{\bR} \equiv\frac{1}{\sqrt{N_s}} {\int_{\text{BZ}}}\frac{d^2q}{(2\pi)^2} e^{i\bq\cdot\bR} \tilde{d}^\dagger_{\hexp\bq}   = \sum_{\bR',\alpha} g_{\bR}(\bR', \alpha) b^\dagger_{\bR',\alpha}.\,\,\,\,
\ee
Here, the function
\be\label{eq:wffuncform}
g_\bR(\bR',\alpha) = {\int_{\text{BZ}}}\frac{d^2q}{(2\pi)^2} \frac{1+e^{i\bq\cdot\mathbf{a}_\alpha}}{\sqrt{N_s\epsilon(\bq)}}{e^{i\bq\cdot(\bR' - \bR)}}
\ee
characterizes the spatial structure of the WOs (Fig. \ref{fig:WFandstates}). It is easily verified from (\ref{eq:wannierdef}) that  the $w_\bR$ are canonical boson operators, i.e.  $[w_\bR, w^\dagger_{\bR'}] = \delta_{\bR,\bR'}$. Filling one boson in each WO gives the Wannier permanent wavefunction \footnote{The exponential decay of $g_{\bR}(i)$ and hence the boson correlations follows from analyticity of $1/\sqrt{\epsilon(\bq)}$ in the BZ.}.

Finally, each individual WO respects the point-group symmetry, as we can explicitly verify.  Translations only map one WO into another and therefore  a product over all the WOs is invariant under the entire space group of the lattice, as is an arbitrary sum of such products, which includes a permanent of the form $\ket{\Psi_W}$. Thus, we conclude that $\ket{\Psi_W}$ represents a featureless Mott insulator at filling $1/3$ \footnote{Note, the permanent wave function is a product state of orthogonal WOs, albeit not a product over disjoint {\it real-space} units. Orthogonality ensures that symmetry of a single WO rules out symmetry-breaking in thermodynamic limit.}.
If we vary the parameters $t_{ij}$ away from the special point, the WOs will change. Nevertheless, as long as the WOs can be chosen to be symmetric and exponentially localized, the corresponding Wannier permanent wave function remains featureless and has exponentially decaying correlations. We thus have a family of Wannier permanents, each characterized by the hopping Hamiltonian $H_0$.

\noindent{\it Parent Hamiltonian.---} We construct a Hamiltonian for which the Wannier permanent wave function $\ket{\Psi_W}$ is an {\em exact} ground state. We do this by first constructing a new noninteracting Hamiltonian, $H_0'$, which flattens the dispersion of the lowest band while leaving its wavefunctions unaffected. Now, the WOs are actually single-particle eigenstates. We tune the chemical potential $\mu$ to be just above the lowest band. This leads to a massive ground state degeneracy, which is lifted by adding appropriate interaction terms that ensure single occupancy of the WOs,  thus selecting $\ket{\Psi_W}$ as the ground state.

We begin with the tight-binding problem $H_0$, and from it obtain the operators $w^\dagger_\bR$. Using these, we construct a {\it different} single-particle Hamiltonian $H_0' = -V\sum_{\bR}w^\dagger_\bR w_\bR - \mu \hat{N}$, where $\hat{N}$ is the total boson number.  Using (\ref{eq:wannierdef}), $H_0'$ can be expressed as a symmetric hopping model on the kagome lattice, whose locality follows from localization of the WOs. Matrix elements of $H_0'$ decay exponentially (identically to $\langle b^\dagger_i b_0\rangle$, Fig.~\ref{fig:WFandstates}), and hopping between sites on non-adjacent hexagons is $\lesssim3\%$ of that between nearest neighbors. It is readily verified that $H_0'$ may be rewritten as
$H_0' = -\sum_\bR[(\mu+V)\hat{n}^{w}_\bR + \mu (\hat{n}^{u_1}_\bR +\hat{n}^{u_2}_\bR)]$
where $\hat{n}^w_\bR = w^\dagger_\bR w_\bR$ represents the occupation of the WO of the lowest band at Bravais lattice site $\bR$, and $\hat{n}_\bR^{u_{1,2}}$ are occupations of the WOs of the two upper bands.

For $-V<\mu<0$ in $H_0'$ only the $n^w_\bR$ are nonzero. There are many ways to fill WOs of the lowest band with bosons,  leading to a huge degeneracy of many-body states. We lift this by adding to $H_0'$ an interaction that penalizes multiple occupancy of a WO. This is accomplished by a term $H_\text{int} = \frac{U}{2}\sum_{\bR}\hat{n}^{w}_\bR(\hat{n}^{w}_\bR-1)$. In terms of WOs, this is the familiar Hubbard interaction stabilizing a Mott phase with fixed occupancy of each WO, but corresponds to a more intricate interaction in terms of the site operators of the  underlying kagome lattice \cite{SupplementaryMaterial}. Like $H_0'$, $H_\text{int}$ is local  in the sense that interactions between sites decay exponentially with their separation. It is easy to show that $\ket{\Psi_W}$  is the ground state of $H_W \equiv H_0' +H_\text{int}$ for $-V<\mu< \min(0, U-V)$ \footnote{Since $n^{u_j}_\bR =0$ for $-V<\mu<0$, and $H_\text{int}$ ensures  $n^w_\bR =1$ for $-V<\mu<U-V$, we obtain $\ket{\Psi_W}$ upon symmetrization.}. It is also clear that all excitations above the ground state are gapped.
 It is evident that $\ket{\Psi_W}$ is the unique state satisfying these properties on the torus, so it is not topologically ordered as it has no ground state degeneracy.
In summary, we have shown $\ket{\Psi_W}$ is the unique gapped and symmetric ground state of  \be\label{eq:HWdef}
H_W = -V\sum_\bR \hat{n}^w_\bR + \frac{U}{2}\sum_\bR \hat{n}^{w}_\bR(\hat{n}^{w}_\bR-1) -\mu \hat{N}
 \ee
for $-V<\mu<\text{min}(0,U-V)$. An equivalent Hamiltonian can be obtained by first flattening the lowest band and then projecting a Hubbard interaction into it \cite{Neupert:2011p1F,Sheng:2011p1F,Tang:2011p1}.
 $H_W$ is intricate and presently challenging to realize experimentally (see below). Numerical study of simpler proximate Hamiltonians that may yield featureless phases is left to future work.

{\noindent{\it Experimental realization.---} A natural experimental setting for Bose-Hubbard physics is in ultracold atomic gases in optical lattice potentials \cite{Greiner:2002p1}.  The kagome geometry was achieved recently in an experimental setup using an optical superlattice, and was characterized by studying properties of an atomic superfluid with such a lattice \cite{Jo:2012p1}.  Mott insulating states within this optical lattice have been recently observed \cite{StamperKurnPrivateCommunication}.}

Methods suitable for identifying an insulating state with fractional filling have been demonstrated for atoms in simple Bravais lattices.  For example, Mott insulating states at various fillings can be identified by observing plateaus in the density of a lattice-trapped gas within an inhomogeneous potential \cite{camp06shell,foll06shell}.  The underlying geometry of the insulating state can be probed via momentum-resolved correlations in the measured atomic density \cite{foll05noise}, a method that can be employed to check against broken-symmetry states.

However, the fine tuning of next- and next-next-nearest neighbor hopping in such lattices may be difficult to achieve.  One approach may be to utilize Feshbach resonances to increase the interaction strength so that a critical ratio of interaction to kinetic energies can be reached already in a very shallow optical lattice, where substantial higher-order tunneling may still occur.  Alternately, returning to the analogy between bosonic tunneling and spin models, longer-range interactions can be achieved and tuned in lattice-trapped gases of polar molecules \cite{buch07polar} or Rydberg atoms.  Although $\ket{\Psi_W}$ is not sign positive, a related sign-positive wavefunction -- potentially with better variational energies -- exists in the same phase, albeit without a corresponding parent Hamiltonian \cite{kptv-wip}.

We note that there are
candidate materials where it is believed that the spin physics is captured by the isotropic kagome lattice Hamiltonian. By exploiting the connection between the Bose-Hubbard model and a spin system in an applied field, the wavefunctions we consider here are  natural candidates for featureless fractional magnetization plateaus in such materials -- particularly since their nontrivial sign structure suggests that they will give good variational energies \cite*{[{Permanent wave functions have applied to magnetization plateaus on the anisotropic triangular lattice by  }] TayMotrunich}.

\noindent{\it Concluding Remarks.---} Although we have focused on a specific example of a non-Bravais Bose-Hubbard model, the kagome lattice at $\frac{1}{3}$ site filling, the path to generalizing our results is clear. For a lattice with a $q$-site unit cell, it is possible to construct a Wannier permanent at filling $1/q$ if we can construct symmetric, exponentially localized WOs for the lowest band.  The first step is to find a tight-binding model whose lowest band is simple, which often requires going beyond nearest-neighbor hopping, as the kagome
teaches us \footnote{The  special simplifying features of the kagome, e.g.  flat bands when
$t_2=t_3=0, t_1$, are
inessential; as long as a simple band exists, the Wannier approach works, and is hence general.}
 In general, whether such a model exists depends on symmetries, as
discussed above. For example, on the $1/2$-filled honeycomb lattice, the two bands that touch in the nearest-neighbor limit  form a two-dimensional irreducible representation; as there is no third trivial band to which the degeneracy may be transferred, the touching cannot be removed without symmetry breaking \cite{SupplementaryMaterial}; and, on non-symmorphic lattices featureless phases are impossible except at special integer fillings \cite{2012arXiv1212.0557P}.

We note in closing that the kagome lattice WO is maximal on the hexagon of sites centered on Bravais lattice site $\bR$ and decays rapidly away from it (Fig.~\ref{fig:WFandstates}.) This suggests that the wavefunction $\ket{\Psi_{\hexp}} = \prod_{\bR} d^\dagger_{\hexp\bR}\ket{0}$ obtained by filling `truncated' orbitals restricted to a hexagon remains in the same phase as $\ket{\Psi_W}$. Verifying this conjecture requires a numerical evaluation of correlations in the state $\ket{\Psi_{\hexp}}$, which is a significant problem in its own right. One can by analogy construct a  featureless insulating wavefunction on the honeycomb lattice, even though the Wannier construction fails \cite{SupplementaryMaterial,kptv-wip}.

\noindent{\it Acknowledgements.---} We thank M. Rypest{\o}l, M. Zaletel and an anonymous referee for comments on the manuscript, which contributed significantly to pedagogy. We acknowledge  support from the Simons Foundation (SAP), the NSF Graduate Fellowship (IK), and the Army Research Office with funding from the DARPA Optical Lattice Emulator program.


\begin{appendix}
\section*{Supplementary Material}

\section{Symmetry Analysis of Band Touchings}
As our approach relies on the ability to construct exponentially-localized Wannier orbitals, it will be important to determine when it is possible to deform a tight-binding model {\it without} breaking lattice symmetry in such a manner as to produce a {\it simple}  band at the requisite filling -- or at the very least, to know in advance when such an effort will prove futile.

In general, there are two ways to produce a simple band: either split degeneracies or just move energy levels around so that the singly degenerate states are all in one band. In this appendix we seek to understand when it is {\it im}possible to produce a simple band while preserving lattice symmetries as this will serve as an important restriction on the approach we take in this paper. Consider a general tight-binding model on a lattice with a specified space group. This is comprised of a point group $G$, combined with translations through vectors  in the Bravais lattice. If the lattice is such that a single center can be chosen about which all the symmetry operations in $G$ may be defined, then the space group is termed {\it symmorphic}; if at least two different centers are required, then the space group is {\it non-symmorphic}. Non-symmorphicity requires that the lattice have screws  (a rotation about an axis combined by a translation along a vector parallel to the axis and not in the Bravais lattice) and/or glides  (a reflection through a plane combined by a translation through a vector in the plane and not in the Bravais lattice), i.e. point-group operations coupled with non-primitive translations\footnote{While symmorphic lattices may have screw and/or glides, these are not required in order to generate the lattice, unlike the nonsymmorphic case}.

We now focus on a tight-binding Hamiltonian on the lattice, and consider the Bloch hamiltonian $h(\bq)$ at crystal momentum $\bq$. For arbitrary $\bq$, no operation in $G$ leaves $\bq$ invariant and therefore is not a symmetry of $h(\bq)$. However, at high-symmetry points in the Brillouin zone, a subgroup of operations in $G$, known as the {\it little group at} $\bq$ and denoted $G_\bq$,  leave $\bq$ invariant or translate it by a reciprocal lattice vector, so that $h(\bq)$ is left invariant, i.e. for $g\in G_\bq$, we have $gh(\bq) g^{-1} = h_\bq$. The energy levels at high-symmetry points can be chosen to be simultaneous eigenstates of both the energy and the generators of $G_\bq$; the levels can be classified by the  irreducible representations of $G_\bq$ that they transform under, and  protection, or lack thereof, of band touchings at high symmetry points may thus be understood in terms of these.  It is to this  understanding that we now turn.

For  a symmorphic space group, all  representations of $G_\bq$ are regular (i.e. non-projective.) In this case, constructing a simple band consistent with the lattice symmetry is impossible if at least one high-symmetry point $\bq$  there are no nondegenerate levels and every degenerate set transforms under an irreducible representation of $G_\bq$ with dimension $\geq2$.  As this statement is somewhat opaque, we illustrate through an example: we will contrast the behavior of the kagome and honeycomb lattices. Both lattices have the symmorphic space group $P6mm$ with a triangular Bravais lattice and point group $D6h$; choosing the center of symmetry to be the center of a hexagon in each case, we can choose the point group generators to be a mirror reflection about the horizontal axis  and  $\pi/3$ rotation  as shown in the figure. The nearest-neighbor tight-binding Hamiltonian for both kagome and honeycomb yields a spectrum with Dirac points at the $K$, $K'$ points of the Brillouin zone where two dispersing bands touch; kagome has in addition a dispersionless flat band at high energy which touches the upper dispersing band at the $\Gamma$ point. To determine whether the Dirac points can be removed without breaking the symmetry, we need to classify the irreducible representations (irreps) of the little group $G_K$ of the $K$-point (or equivalently, of the $K'$ point). We shall discuss the case of the $K$ point; our arguments apply {\it mutatis mutandis} to the $K'$ point as well. Since a rotation by $\pi/3$ maps $K$ to $K'$, it is clear that $R_{\pi/3}\not\in G_K$; only rotations by $2\pi/3$ leave $K$ invariant, so only the generator of threefold symmetry $R_{2\pi/3} = R_{\pi/3}^2$ is in the little group. The other little group generator is the mirror reflection $\sigma_2$ -- which also leaves $K$ invariant -- and together these generate  $G_K\cong D_{3h}$. For the kagome lattice, it is evident that the rotation cyclically permutes the sublattice indices $A$, $B$, and $C$   (corresponding sites denoted by red, blue and green dots respectively in all the figures.) The mirror reflection interchanges $B$ and $C$ while leaving $A$ invariant. Since both these operations may be written (as is evident from our choice of unit cell labeling) as a mapping $\bR\rightarrow g\bR$, accompanied by a permutation of elements in the unit cell we may represent the generators of  $G_K$ by the matrices
\be
R_{2\pi/3} = \left(\begin{array}{ccc} 0&1&0\\0&0&1\\1&0&0\end{array}\right), \sigma_2 = \left(\begin{array}{ccc} 1&0&0\\0&0&1\\0&1&0\end{array}\right)
\ee
Since both these leave the vector $(1,1,1)$ invariant, it is clear that the representation they generate is reducible (which we could have deduced from the fact that $D_{3h}$ has no 3-dimensional irreps) and decomposes into the direct sum of a one-dimensional and a two-dimensional irrep.  We can show that eigenstates from the pair of bands that touch at $K$ in the nearest-neighbor limit form a two-dimensional irrep of $G_K$,while the flat band eigenstates must form a one-dimensional irrep. As result, it is possible to construct a simple band on the kagome lattice as we demonstrate explicitly in the paper. To do this,  we add higher-neighbor hopping terms and tune their strength until a threefold degeneracy occurs at $K$; once this degeneracy point is crossed, the upper two bands form the  two-dimensional representation and are degenerate at $K$, while the lower band transforms under the trivial representation at the $K$ point. The band touching in the honeycomb lattice on the other hand is protected by symmetry: since there are no non degenerate bands at $K$ that form a trivial representation of $G_K$, there is no way to remove the band touching by a procedure analogous to the kagome example.

Turning next to the honeycomb lattice, we note that the mirror reflection about the axis interchanges $A$ and $B$ and maps $\bR\rightarrow \sigma_2\bR$. However, the $2\pi/3$ rotation necessarily mixes sites from different unit cells, unlike in the kagome case: we can show that this operation takes $\bR\rightarrow{R}_{2\pi/3}\bR$, but also maps (in an obvious notation) $A(\bR)\rightarrow A(\bR+\mathbf{a}_1+\mathbf{a}_2)$, $B(\bR)\rightarrow B(\bR -\mathbf{a}_1-\mathbf{a}_2)$. In Fourier space and at the $K$-point, it is clear that the representation matrices corresponding to these generators take the form
\be
R_{2\pi/3} = \left(\begin{array}{cc} e^{i2\pi/3}&0\\0&e^{-i2\pi/3}\end{array}\right), \sigma_2 = \left(\begin{array}{cc} 0&1\\1&0\end{array}\right)
\ee
It is readily verified that these form a two-dimensional irrep, and thus the Dirac point degeneracy is protected by symmetry in the honeycomb case.
\begin{figure}
\includegraphics[width=0.8\columnwidth]{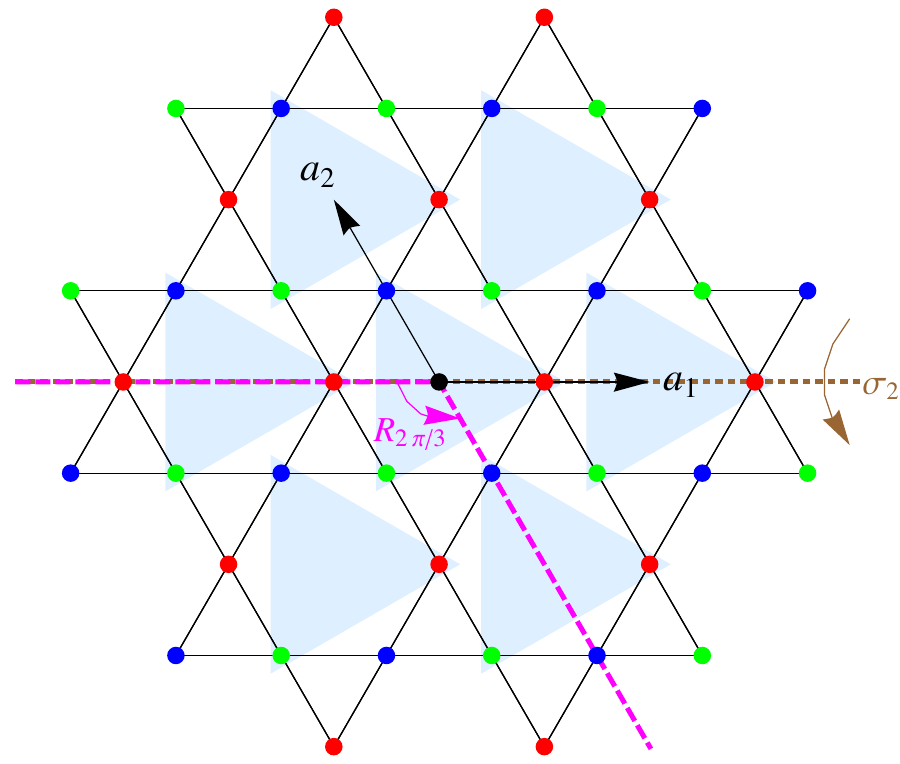}
\caption{\label{fig:kagome} Kagome lattice. The center of symmetry is shown, and our choice of sites in a unit cell is shaded, and the basis vectors of the triangular Bravais lattice are shown. The little group generators referred to in the text are also depicted: (i) rotations $R_{2\pi/3}$ by 120 degrees about the center; and (ii) reflections $\sigma_2$ about the mirror line.}
\end{figure}

\begin{figure}
\includegraphics[width=0.8\columnwidth]{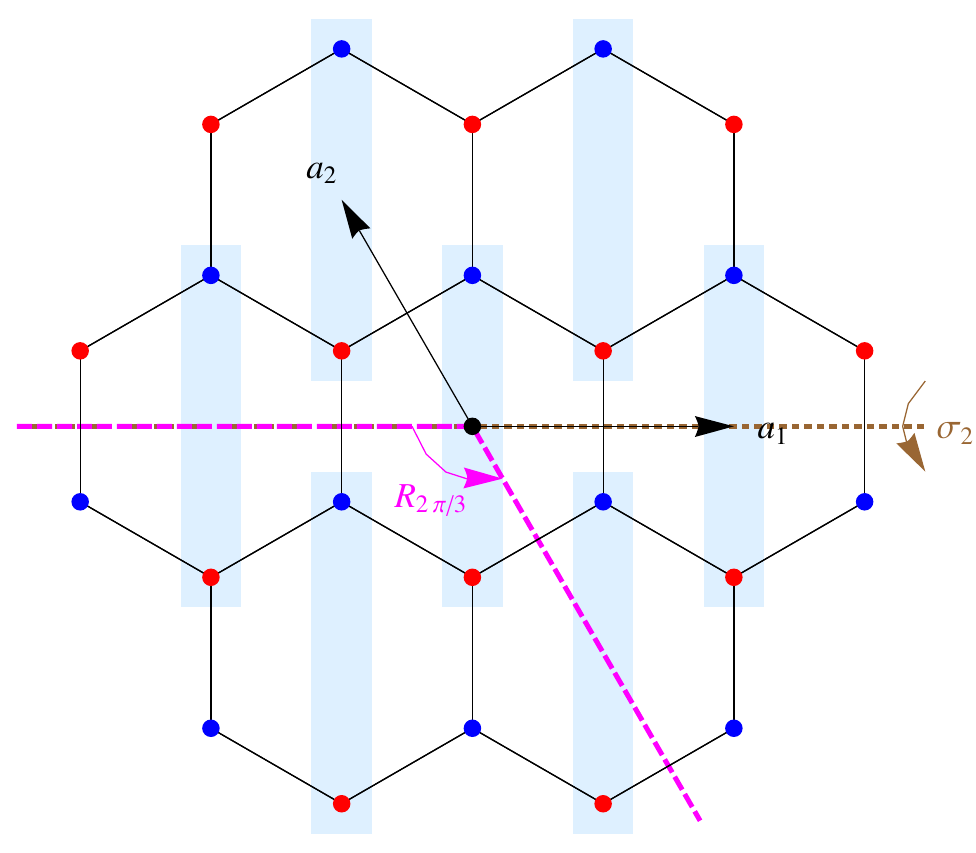}
\caption{\label{fig:honeycomb} Honeycomb lattice. We use the same conventions and notations as Fig.~\ref{fig:kagome}.}
\end{figure}

\begin{figure}
\includegraphics[width=0.8\columnwidth]{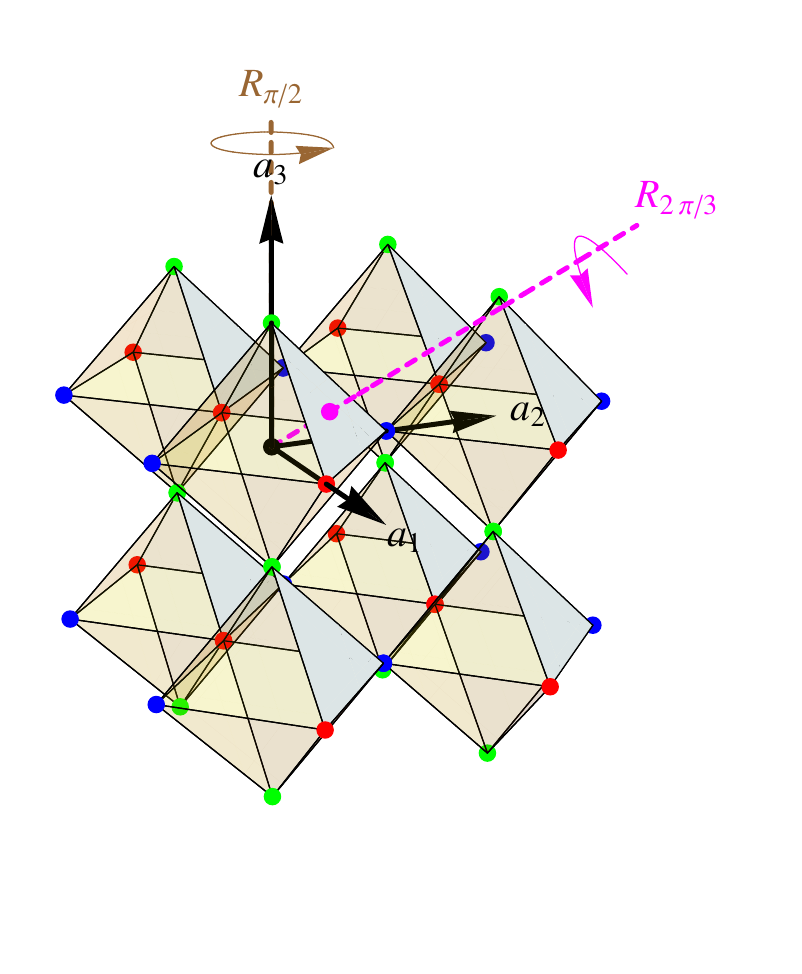}
\caption{\label{fig:perovskite} Cubic perovskite structure viewed as a lattice of corner sharing octahedra. We show only the symmetries required to prove that $G_{(\pi,\pi,\pi)}$ has a three-dimensional irrep.}
\end{figure}

A three-dimensional example of a symmorphic space group with a protected band touching is furnished by the cubic perovskite structure, which is a lattice of corner-sharing octahedra; alternatively we may obtain this by placing sites at the bond mid-points of a simple cubic lattice. We may then verify that the energy bands at $R=(\pi,\pi,\pi)$  transform under a three-dimensional irrep of the little group, as follows. Rather than construct the entire little group $G_R$, we will content ourselves merely with constructing a pair of elements and use them to show that the three-dimensional representation is irreducible. The first of these is the rotation $R_{2\pi/3}$ about the $(1,1,1)$ direction (see Fig.~\ref{fig:perovskite}). This simply sends $\bR \rightarrow R_{2\pi/3}\bR$, and cyclically permutes $A$,$B$, and $C$; the second is the rotation by $R_{\pi/2}$ about the $z$-axis; this maps $\bR\rightarrow R_{\pi/2}\bR$ and leaves $C$ invariant, while taking $A(\bR) \rightarrow B(\bR)$, $B(\bR)\rightarrow A(\bR - \mathbf{a_1})$. The corresponding representation matrices are (at the $R$-point)
\be
R_{2\pi/3} = \left(\begin{array}{ccc} 0&1&0\\0&0&1\\1&0&0\end{array}\right), R_{\pi/2} = \left(\begin{array}{ccc} 0&-1&0\\1&0&0\\0&0&1\end{array}\right)
\ee
which share no common eigenvectors. Any representation of $G_R$ must contain these two symmetry operations and thus cannot be reducible, and therefore the threefold degeneracy at $R$ is protected by symmetry.

For a non-symmorphic space group, representations of $G_\bq$ for $\bq$ on the surface of the Brillouin zone are in general projective. As a result, if $G_\bq$ contains commuting elements that fail to commute in the projective representation, symmetry requires that every level at $\bq$ be degenerate. For instance, we may show that if $G_\bq$  contains inversion plus a nonsymmorphic generator, then {\it every} energy level is at least twofold degenerate at $\bq$ and therefore a simple band cannot be constructed without breaking lattice symmetry. An example of this is the three-dimensional space group $Fd3m$, which characterizes the diamond and pyrochlore lattices: we can show that the little group $G_X$ of the $X$-point contains inversion and a four-fold screw rotation, which commute. As the screw is nonsymmorphic, the representations of $G_X$  are projective and therefore every band at $X$ is (at least) twofold degenerate, precluding the existence of any simple bands. For a  lucid explanation of such symmetry-required band stickings, we refer the reader to \cite{Konig:1997p1, Konig:1999p1, Mermin:1992p1}. 

\section{Parent Hamiltonian}
In the main text, we introduced a parent Hamiltonian for the Wannier permanent. Here we clarify a few of its essential properties.

The first subtlety is that  we rewrote the noninteracting portion of the Hamiltonian, $H_0' = -V\sum_\bR w^\dagger_\bR w_\bR - \mu \hat{N}$ in terms of Wannier orbitals of the upper bands, $H_0'  = -\sum_\bR[(\mu+V)\hat{n}^{w}_\bR + \mu (\hat{n}^{u_1}_\bR +\hat{n}^{u_2}_\bR)]$. We observe that the terms corresponding to occupying WOs of the upper bands are contributed only by the chemical potential term, and are essential since the chemical potential couples to the {\it total} particle number $\hat{N}$, which is impossible to expand purely in terms of the lowest band. It is easily verified that $[\hat{n}^W_\bR, \hat{n}^{u_{1,2}}_\bR] = 0$, and also $[\hat{n}^W_\bR, \hat{n}^W_{\bR'}] =[\hat{n}^W_\bR, \hat{n}^{u_{1,2}}_{\bR'}] =0$. Thus, for $-V<\mu<0$, it is possible to set the occupation of the upper bands to zero, since the remaining terms in the Hamiltonian $H_W = H_0' +H_\text{int}$ do not mix the lowest band with the upper two bands. Also, it should be clear that the symmetry and localization properties of the upper two bands are irrelevant since they enter $H_0'$ in a manner that manifestly preserves the symmetry and is local, since all terms proportional to $\mu$ simply combine to give the total particle number $\hat{N} = \sum_i b^\dagger_i b_i$.

A second point is that  the final Hamiltonian $H_W$ contains terms that look `onsite' on the triangular Bravais lattice, but in fact are quite intricate on the kagome. The explicit form of $H_W$ can be seen by rewriting the Wannier orbital boson operators $w^\dagger_\bR$ in terms of the original kagome lattice boson operators $b^\dagger_i$. From Eq.(3)  of the main text,
\be
w^\dagger_\bR = \sum_{\bR',\alpha} g_\bR(\bR',\alpha) b^\dagger_{\bR',\alpha} =  \sum_{i} g_\bR(i) b^\dagger_i 
\ee
where we have reintroduced the compact notation $i \equiv(\bR',\alpha)$ for sites on the kagome lattice. Using this, we obtain
\be
H_W &=&  \sum_{ij} \left(-V+\frac{U}{2}\right) f^{(2)}_{ij} b^\dagger_i b_j  - \mu\sum_i b^\dagger_i b_i \nonumber\\& & - \frac{U}{2} \sum_{ijkl} f^{(4)}_{ijkl} b^\dagger_i b^\dagger_j b_k b_l   \ee
where the interaction `form factors'
\be
 f^{(2)}_{ij}  &=& \sum_{\bR} g_{\bR}(i)g^*_\bR(j) \nonumber\\
 f^{(4)}_{ijkl} &=& \sum_{\bR} g_{\bR}(i)g_\bR(j) g^*_{\bR}(k)g^*_\bR(l)
\ee
are computed from the Wannier orbital wavefunction. Owing to the exponential localization of $g_\bR$, they decay exponentially with separation between any pair of sites.

In discussion above we specialized to the kagome, but it should be evident that with appropriate changes can be adapted to any simple band. 
\end{appendix}
\bibliography{mottbib}
\end{document}